\documentclass[12pt]{iopart}
\usepackage{amsfonts}
\usepackage{setstack}
\usepackage{graphicx,graphics}

\begin{document}

\title{The effect of six-point one-particle reducible local interactions in
the dual fermion approach }
\author{A. A. Katanin \\
%EndAName
Institute of Metal Physics, 620990, Ekaterinburg, Russia\\
Ural Federal University, 620002 Ekaterinburg, Russia}

\begin{abstract}
We formulate the dual fermion approach for strongly correlated electronic
systems in terms of the lattice and dual effective interactions, obtained by
using the covariation splitting formula. This allows us to consider the
effect of six-point one-particle reducible interactions, which are usually
neglected by the dual fermion approach. We show that the consideration of
one-particle reducible six-point (as well as higher order) vertices is
crucially important for the diagrammatic consistency of this approach. In
particular, the relation between the dual and lattice self-energy, derived
in the dual fermion approach, implicitly accounts for the effect of the
diagrams, containing 6-point and higher order local one-particle reducible
vertices, and should be applied with caution, if these vertices are
neglected. Apart from that, the treatment of the self-energy feedback is
also modified by 6-point and higher order vertices; these vertices 
%treatment of the 6-point and higher order local one-particle reducible
%vertices in the dual fermion approach is
are also important to account for some non-local corrections to the lattice
self-energy, which have the same order in the local 4-point vertices, as the
diagrams usually considered in the approach. These observations enlighten 
%a question of applicability of the approximation,
an importance of 6-point and higher order vertices in the dual fermion
approach, and call for development of new schemes of treatment of non-local
fluctuations, which are based on one-particle irreducible quantities.
\end{abstract}

\maketitle

%\address{$^a$ Max-Planck-Institut f\"ur Festk\"orperforschung, D-70569,\\
%Stuttgart\\
%$^b$ Institute of Metal Physics, 620219 Ekaterinburg, Russia}

\section{Introduction}

Strongly correlated electron systems became one of the touchstone of modern
physics. They demonstrate a variety of phenomena: magnetism,
(unconventional) superconductivity, \textquotedblleft
colossal\textquotedblright\ magnetoresistance, and quantum critical
behavior. The dynamical mean-field theory (DMFT)\cite{DMFT,DMFT2}\ allowed
to describe accurately the Mott-Hubbard metal-insulator transition.\cite{MH}%
. DMFT becomes exact in the limit of high spatial dimensions ($d\rightarrow
\infty $) and accounts for an important local part of electronic
correlations. Real physical systems are however one-, two-, or
three-dimensional. Therefore, nonlocal correlations, which are neglected in
DMFT, may be important. Recently, a progress to go beyond DMFT through
cluster extensions\cite%
{clusterDMFT,clusterDMFT1,clusterDMFT2,clusterDMFT3,clusterDMFT4} was
achieved. These correlations are however necessarily short-range in nature
due to numerical limitations of the cluster size\cite{clusterDMFTFlex}.

This limitation motivated developing the diagrammatic extensions of the
dynamical mean-field theory. The dynamical vertex approximation (D$\Gamma $%
A) was introduced in Refs. \cite{DGA1a,DGA1b,DGA1c,DGA1d,DGA2,Kusunose}.
Starting from the local particle-hole irreducible vertex, this approximation
sums ladder diagrams for the vertex in the particle-hole channel, where
particle-hole irreducible vertices assumed to be local, but the effect of
the non-locality of the Green functions is considered. Alternative dual
fermion (DF) approach was proposed in Refs. \cite{DF1,DF2,DF3,DF4}, which
splits the degrees of freedom into the local ones, treated within DMFT, and
the non-local (dual) degrees, considered perturbatively, with a possibility
of summation of infinite series of diagrams for dual fermions \cite%
{DF4,Jarrell}.

Although both abovementioned approaches use 4-point local vertex as an
effective interaction between fermionic degrees of freedom (lattice fermions
in case of $D\Gamma A$ and dual fermions in the DF approach), they make in
fact very different assumptions on the neglect of higher-order local
vertices. Indeed, $D\Gamma A$ operates with one-particle irreducible (1PI)
vertices, and neglects six-point and higher order 1PI local vertices. At the
same time, DF representation does not use the assumption of the one-particle
irreducibility; in particular its formulation in Refs. \cite{DF1,DF2,DF3,DF4}
neglects one-particle \textit{reducible} six-point and higher vertices.

This difference appears to be important for analysing diagrammatic
consistency of the abovediscussed approaches. While the dynamic vertex
approximation is based on the diagrammatic approach, formulated in terms of
the original lattice degrees of freedom, the diagrammatic consistency of the
dual fermion approach (in terms of the same lattice degrees of freedom) has
to be verified. In the present paper we show that the inculision of the
one-particle reducible six-point (and more generally, higher vertices) into
the DF approach appears to be necessary to make it diagrammatically
consistent.

\section{The model, dynamical mean-field theory, and the dual fermion
approach}

\subsection{The model and dynamical mean-field theory}

We consider general one-band model of fermions, interacting via local
interaction $H_{\mathrm{int}}[\widehat{c}_{i\sigma },\widehat{c}_{i\sigma
}^{+}]$ 
\begin{equation}
H=\sum\limits_{\mathbf{k},\sigma }\varepsilon _{\mathbf{k},\sigma }\widehat{c%
}_{\mathbf{k},\sigma }^{+}\widehat{c}_{\mathbf{k},\sigma
}+\sum\limits_{i,\sigma }H_{\mathrm{int}}[\widehat{c}_{i\sigma },\widehat{c}%
_{i\sigma }^{+}],  \label{H}
\end{equation}%
where $\widehat{c}_{i\sigma },\widehat{c}_{i\sigma }^{+}$ are the fermionic
operators, and $\widehat{c}_{\mathbf{k},\sigma },\widehat{c}_{\mathbf{k}%
,\sigma }^{+}$ are the corresponding Fourier transformed objects, $\sigma
=\uparrow ,\downarrow $ corresponds to a spin index. The model is
characterized by the generating functional%
\begin{eqnarray}
Z[\eta ,\eta ^{+}] &=&\int d[c,c^{+}]\exp \left\{ -\mathcal{S}[c,c^{+}]+\eta
^{+}c+c^{+}\eta \right\}  \label{gen} \\
\mathcal{S}[c,c^{+}] &=&\int d\tau \left\{ \sum\limits_{i,\sigma }c_{i\sigma
}^{\dagger }(\tau )\frac{\partial }{\partial \tau }c_{i\sigma }(\tau
)+H[c,c^{+}]\right\}
\end{eqnarray}%
where $c_{i\sigma },c_{i\sigma }^{+},\eta _{i\sigma },\eta _{i\sigma }^{+}$
are the Grassman fields, the fields $\eta _{i\sigma },\eta _{i\sigma }^{+}$
correspond to source terms, $\tau \in \lbrack 0,\beta =1/T]$ is the
imaginary time. The dynamical mean-field theory corresponds to considering
the effective single-site problem with the action%
\begin{eqnarray}
\mathcal{S}_{\mathrm{DMFT}}[c,c^{+}] &=&\sum_{i,\sigma }\int d\tau \left\{ -%
\frac{1}{\beta ^{2}}\int d\tau ^{\prime }\zeta ^{-1}(\tau -\tau ^{\prime
})c_{i\sigma }^{\dagger }(\tau )c_{i\sigma }(\tau ^{\prime })\right.
\label{LDMFT} \\
&& +H_{\mathrm{int}}[c_{i\sigma },c_{i\sigma }^{+}]\Bigg\} ,  \nonumber
\end{eqnarray}%
where the "Weiss field" function $\zeta (\tau )$ and its Fourier transform $%
\zeta (i\nu _{n})$ has to be determined self-consistently from the condition%
\begin{equation}
G_{\mathrm{loc}}(i\nu _{n})\equiv \frac{1}{\zeta ^{-1}(i\nu _{n})-\Sigma _{%
\mathrm{loc}}(i\nu _{n})}=\sum\limits_{\mathbf{k}}G(\mathbf{k},i\nu _{n})
\label{sc}
\end{equation}%
where%
\begin{equation}
G(\mathbf{k},i\nu _{n})\equiv G_{k}=\left[ G_{0k}^{-1}-\Sigma _{\mathrm{loc}%
}(i\nu _{n})\right] ^{-1}  \label{Glattice}
\end{equation}%
$G_{0k}^{-1}=i\nu _{n}-\varepsilon _{\mathbf{k}}\,$is the lattice
noninteracting Green function (we use the $4$-vector notation $k=(\mathbf{k}%
,i\nu _{n})$) and $\Sigma _{\mathrm{loc}}(i\nu _{n})$ is the self-energy of
the impurity problem (\ref{LDMFT}), which is in practice obtained within one
of the impurity solvers: exact diagonalization, quantum Monte-Carlo, etc.

\subsection{The formulation of the dual fermion approach by means of
covariation splitting formula}

The dual fermion approach of Refs. \cite{DF1,DF2,DF3,DF4} can be
conveniently formulated in terms of an effective interaction of the lattice
theory (see, e.g. Ref. \cite{Salmhofer}) 
\begin{eqnarray}
\mathcal{V}[\eta ,\eta ^{+}] &:&=-\ln \int d[c,c^{+}]\exp \left\{
\sum\limits_{k,\sigma }G_{0k}^{-1}\left( c_{k\sigma }^{+}+\eta _{k\sigma
}^{+})(c_{k\sigma }+\eta _{k\sigma }\right) \right.  \nonumber \\
&&\left. -\sum\limits_{i,\sigma }\int d\tau H_{\mathrm{int}}[c_{i\sigma
},c_{i\sigma }^{+}]\right\}  \nonumber \\
&=&-\ln Z[G_{0k}^{-1}\eta _{k\sigma },G_{0k}^{-1}\eta _{k\sigma }^{+}]-\eta
_{k\sigma }^{+}G_{0k}^{-1}\eta _{k\sigma }.  \label{Vs}
\end{eqnarray}%
Expansion of the effective interaction $\mathcal{V}[\eta ,\eta ^{+}]$ in
source fields generates connected (in general, one-particle reducible) Green
functions, amputated by the non-interacting Green functions of the lattice
theory $G_{0k}$. The relation between one-particle reducible and 1PI
counterparts of the vertices can be involved. In particular, the
(one-particle irreducible) self-energy $\Sigma _{k}$ of the lattice problem
(i.e. the 1PI 2-point vertex function) can be extracted from the two-point
connected vertex function $V^{(2)}$ via the relation $V_{k}^{(2)}=\Sigma
_{k}/(1-G_{0k}\Sigma _{k})$.

To split the local and non-local degrees of freedom in the effective
interaction (\ref{Vs}) we use the covariation splitting formula \cite%
{Salmhofer}, which is based on the identity%
\begin{eqnarray}
&&\sum\limits_{k,\sigma }C_{k}^{-1}c_{k\sigma }^{+}c_{k\sigma }=\ln \int d[%
\widetilde{c},\widetilde{c}^{+}]\exp \left\{ \sum\limits_{k,\sigma }%
\widetilde{c}_{k,\sigma }^{+}B_{k}^{-1}\widetilde{c}_{k,\sigma }\right.
\label{Id} \\
&&\ \ \ \ \ \ \ \ \ \ \ \ \ \ \ \ \ \ \ \ \ \ \ \ \ \ \ \ \ \ \ \ \ \left.
+\sum\limits_{k\sigma }A_{k}^{-1}\left( \widetilde{c}_{k,\sigma
}^{+}+c_{k,\sigma }^{+})(\widetilde{c}_{k,\sigma }+c_{k,\sigma }\right)
\right\}  \nonumber
\end{eqnarray}%
with $A_{k}+B_{k}=C_{k}$; Eq. (\ref{Id}) can be proven by integrating over
the $\widetilde{c},\widetilde{c}^{+}$ fields. For $A_{k}=\zeta (i\nu _{n}),$ 
$C_{k}=G_{0k}$ this implies%
\begin{equation}
\mathcal{V}[\eta ,\eta ^{+}]=-\ln \int d[\widetilde{c},\widetilde{c}%
^{+}]\exp \left\{ \sum\limits_{k,\sigma }\widetilde{c}_{k,\sigma }^{+}%
\widetilde{G}_{0k}^{-1}\widetilde{c}_{k,\sigma }-\mathcal{V}_{\mathrm{DMFT}%
}[\eta ^{+}+\widetilde{c}^{+},\eta +\widetilde{c}]\right\}  \label{Vs1}
\end{equation}%
where $\mathcal{V}_{\mathrm{DMFT}}[\eta ,\eta ^{+}]$\textit{\ }is\textit{\ }%
an\textit{\ }effective potential\textit{\ }of the dynamical mean-field
theory, defined by%
\begin{eqnarray}
e^{-\mathcal{V}_{\mathrm{DMFT}}[\eta ,\eta ^{+}]} &=&\int d[c,c^{+}]\exp
\left\{ -\sum\limits_{i,\sigma }\int d\tau H_{\mathrm{int}}[c_{i\sigma
},c_{i\sigma }^{+}]\right.  \nonumber \\
&&\left. +\sum\limits_{k,\sigma }\zeta ^{-1}(i\nu _{n})\left( c_{k,\sigma
}^{+}+\eta _{k,\sigma }^{+})(c_{k,\sigma }+\eta _{k,\sigma }\right) \right\}
,
\end{eqnarray}%
$\widetilde{G}_{0k}=G_{0k}-\zeta (i\nu _{n})$ is the bare Green function of
the non-local degrees of freedom. Similarly to $\mathcal{V}[\eta ,\eta ^{+}]$%
, the functional $\mathcal{V}_{\mathrm{DMFT}}[\eta ,\eta ^{+}]$ generates
connected vertices (which are in general one-particle reducible), amputated
by the bath Green function $\zeta (i\nu _{n})$.

To simplify Eq. (\ref{Vs1}), we perform a shift $\widetilde{c}_{k\sigma
}\rightarrow \widetilde{c}_{k\sigma }-\eta _{k\sigma },$ such that%
\begin{equation}
\mathcal{V}[\eta ,\eta ^{+}]=-\ln \int d[\widetilde{c},\widetilde{c}%
^{+}]e^{\sum\limits_{k,\sigma }\left( \widetilde{c}_{k,\sigma }^{+}-\eta
_{k,\sigma }^{+}\right) \widetilde{G}_{0k}^{-1}\left( \widetilde{c}%
_{k,\sigma }-\eta _{k,\sigma }\right) -\mathcal{V}_{\mathrm{DMFT}}[%
\widetilde{c},\widetilde{c}^{+}]}.
\end{equation}%
To arrive at the standard dual fermion approach\cite{DF1,DF2,DF3,DF4} we
consider an expansion of $\mathcal{V}_{\mathrm{DMFT}}[\widetilde{c},%
\widetilde{c}^{+}]$ in fields%
\begin{eqnarray}
\mathcal{V}_{\mathrm{DMFT}}[\widetilde{c},\widetilde{c}^{+}]
&=&\sum\limits_{k,\sigma }\widetilde{c}_{k,\sigma }^{+}\frac{\Sigma _{%
\mathrm{loc}}(i\nu _{n})}{1-\zeta (i\nu _{n})\Sigma _{\mathrm{loc}}(i\nu
_{n})}\widetilde{c}_{k,\sigma }+\widetilde{\mathcal{V}}_{\mathrm{DMFT}}[%
\widetilde{c},\widetilde{c}^{+}]  \nonumber \\
\widetilde{\mathcal{V}}_{\mathrm{DMFT}}[\widetilde{c},\widetilde{c}^{+}] &=&%
\frac{1}{2}\widetilde{\Gamma }_{\mathrm{loc}}\circ (\widetilde{c}%
_{k_{1\sigma }}^{+}\widetilde{c}_{k_{2\sigma }})(\widetilde{c}_{k_{3}\sigma
^{\prime }}^{+}\widetilde{c}_{k_{4}\sigma ^{\prime }})  \label{VDMFT} \\
&&+\frac{1}{6}\widetilde{\Gamma }_{\mathrm{loc}}^{(6)}\circ (\widetilde{c}%
_{k_{1\sigma }}^{+}\widetilde{c}_{k_{2\sigma }})(\widetilde{c}_{k_{3}\sigma
^{\prime }}^{+}\widetilde{c}_{k_{4}\sigma ^{\prime }})(\widetilde{c}%
_{k_{5}\sigma ^{\prime \prime }}^{+}\widetilde{c}_{k_{6}\sigma ^{\prime
\prime }})+...  \nonumber
\end{eqnarray}%
where $\widetilde{\Gamma }_{\mathrm{loc}}$ and $\widetilde{\Gamma }_{\mathrm{%
loc}}^{(6)}$ are the connected 4- and 6- point local vertices, amputated
with the bare Green functions $\zeta ,$ e.g. 
\begin{eqnarray}
\widetilde{\Gamma }_{\mathrm{loc}}^{\sigma \sigma ^{\prime }}(i\nu
_{1}..i\nu _{3}) &=&(1+\delta _{\sigma \sigma ^{\prime
}})^{-1}\prod\nolimits_{i=1}^{4}\zeta ^{-1}(i\nu _{i})  \label{Gamma_loc} \\
&&\times \left[ G_{\mathrm{loc,}\sigma \sigma ^{\prime }}^{(4)}(i\nu
_{1}..i\nu _{3})-G_{\mathrm{loc}}(i\nu _{1})G_{\mathrm{loc}}(i\nu
_{2})(\delta _{\nu _{1}\nu _{3}}-\delta _{\sigma \sigma ^{\prime }}\delta
_{\nu _{2}\nu _{3}})\right] ,  \nonumber
\end{eqnarray}%
and $\circ $ stands for summation over momenta- frequency- and spin indices
fulfilling the conservation laws, $G_{\mathrm{loc}}^{(4)}$ is the
two-particle local Green function, which can be obtained via the solution of
the impurity problem\cite{DGA1a,Toschi}. We therefore obtain%
\begin{eqnarray}
\mathcal{V}[\eta ,\eta ^{+}] &=&-\sum\limits_{k,\sigma }\eta _{k,\sigma }^{+}%
\widetilde{G}_{0k}^{-1}\eta _{k,\sigma }  \label{Vdual} \\
&&-\ln \int d[\widetilde{c},\widetilde{c}^{+}]e^{\sum\limits_{k,\sigma }%
\left[ \widetilde{c}_{k,\sigma }^{+}(\widetilde{G}_{k}^{\prime })^{-1}%
\widetilde{c}_{k,\sigma }-\widetilde{c}_{k,\sigma }^{+}\widetilde{G}%
_{0k}^{-1}\eta _{k,\sigma }-\eta _{k,\sigma }^{+}\widetilde{G}_{0k}^{-1}%
\widetilde{c}_{k,\sigma }\right] -\widetilde{\mathcal{V}}_{\mathrm{DMFT}}[%
\widetilde{c},\widetilde{c}^{+}]},  \nonumber
\end{eqnarray}%
where%
\begin{equation}
\widetilde{G}_{k}^{\prime }=\widetilde{G}_{0k}\frac{1-\zeta (i\nu
_{n})\Sigma _{\mathrm{loc}}(i\nu _{n})}{1-G_{0k}\Sigma _{\mathrm{loc}}(i\nu
_{n})}=\left[ 1-\zeta (i\nu _{n})\Sigma _{\mathrm{loc}}(i\nu _{n})\right]
^{2}\widetilde{G}_{k},
\end{equation}%
and $\widetilde{G}_{k}=G_{k}-G_{\mathrm{loc}}(i\omega _{n}).$ Rescaling the
fields of integration to exclude extra factor $(1-\zeta \Sigma _{\mathrm{loc}%
})^{2}$ and introducing the `dual' source field 
\begin{equation}
\widehat{\eta }_{k\sigma }=\eta _{k\sigma }/\left[ 1-\Sigma _{\mathrm{loc}%
}(i\nu _{n})G_{0k}\right] ,  \label{eta}
\end{equation}%
we obtain the effective interaction of the lattice theory in the form 
\begin{eqnarray}
\mathcal{V}[\eta ,\eta ^{+}] &=&\widehat{\mathcal{V}}[\widehat{\eta },%
\widehat{\eta }^{+}]-\sum\limits_{k,\sigma }\eta _{k,\sigma }^{+}\left\{ 
\widetilde{G}_{0k}^{-1}-\frac{1}{[1-\Sigma _{\mathrm{loc}}(i\nu
_{n})G_{0k}]^{2}}\widetilde{G}_{k}^{-1}\right\} \eta _{k,\sigma }
\label{Vdtol} \\
&=&\widehat{\mathcal{V}}[\widehat{\eta },\widehat{\eta }^{+}]+\sum_{k,\sigma
}\eta _{k,\sigma }^{+}\frac{\Sigma _{\mathrm{loc}}(i\nu _{n})}{1-\Sigma _{%
\mathrm{loc}}(i\nu _{n})G_{0k}}\eta _{k,\sigma },
\end{eqnarray}%
where%
\begin{equation}
\widehat{\mathcal{V}}[\widehat{\eta },\widehat{\eta }^{+}]=-\ln \int D[%
\widetilde{c},\widetilde{c}^{+}]e^{\sum\limits_{k,\sigma }\widetilde{G}%
_{k}^{-1}\left( \widetilde{c}_{k,\sigma }^{+}-\widehat{\eta }_{k,\sigma
}^{+}\right) \left( \widetilde{c}_{k,\sigma }-\widehat{\eta }_{k,\sigma
}\right) -\widehat{\mathcal{V}}_{\mathrm{DMFT}}[\widetilde{c}^{+},\widetilde{%
c}]}  \label{Vez1}
\end{equation}%
is the effective interaction of dual fermions. According to the Eq. (\ref%
{VDMFT}), the expansion of $\widehat{\mathcal{V}}_{\mathrm{DMFT}}[\widetilde{%
c}^{+},\widetilde{c}]$ in fields reads 
\begin{eqnarray}
\widehat{\mathcal{V}}_{\mathrm{DMFT}}[\widetilde{c}^{+},\widetilde{c}] &=&%
\frac{1}{2}\Gamma _{\mathrm{loc}}\circ (\widetilde{c}_{k_{1},\sigma }^{+}%
\widetilde{c}_{k_{3},\sigma })(\widetilde{c}_{k_{2},\sigma ^{\prime }}^{+}%
\widetilde{c}_{k_{4},\sigma ^{\prime }})  \label{VDMFT1} \\
&&+\frac{1}{6}\Gamma _{\mathrm{loc}}^{(6)}\circ (\widetilde{c}_{k_{1\sigma
}}^{+}\widetilde{c}_{k_{2\sigma }})(\widetilde{c}_{k_{3}\sigma ^{\prime
}}^{+}\widetilde{c}_{k_{4}\sigma ^{\prime }})(\widetilde{c}_{k_{5}\sigma
^{\prime \prime }}^{+}\widetilde{c}_{k_{6}\sigma ^{\prime \prime }})+... 
\nonumber
\end{eqnarray}%
where $\Gamma _{\mathrm{loc}}=\widetilde{\Gamma }_{\mathrm{loc}%
}\prod\nolimits_{i=1}^{4}(1-\zeta \Sigma _{\mathrm{loc}})_{i\nu _{i}}$ and $%
\Gamma _{\mathrm{loc}}^{(6)}=\widetilde{\Gamma }_{\mathrm{loc}%
}^{(6)}\prod\nolimits_{i=1}^{6}(1-\zeta \Sigma _{\mathrm{loc}})_{i\nu _{i}}$
are the connected $4$- and $6$-point vertices, amputated with the local
Green functions $G_{\mathrm{loc}}.$ For the four-point vertex $\Gamma _{%
\mathrm{loc}}$ the requirement of connectivity and amputation with the full
local Green functions implies one-particle irreducibility. However, the
higher-order vertices, e.g. $\Gamma _{\mathrm{loc}}^{(6)}$ remain
one-particle reducible.

In the described approach the relation between the self-energies $\Sigma _{%
\mathrm{d}}(k)$ of the dual fermions $\widetilde{c}_{k}$ and $\Sigma _{k}$
of the lattice fermions $c_{k}$ is easily obtained by equating the
corresponding two-point vertices in Eq. (\ref{Vdtol}):%
\begin{eqnarray}
\frac{\Sigma _{k}}{1-\Sigma _{k}G_{0k}} &=&\frac{\Sigma _{\mathrm{loc}}(i\nu
_{n})}{1-\Sigma _{\mathrm{loc}}(i\nu _{n})G_{0k}}  \nonumber \\
&&+\frac{\Sigma _{\mathrm{d}}(k)}{1-\Sigma _{\mathrm{d}}(k)\left[ G_{k}-G_{%
\mathrm{loc}}(i\nu _{n})\right] }\frac{1}{\left[ 1-\Sigma _{\mathrm{loc}%
}(i\nu _{n})G_{0k}\right] ^{2}},
\end{eqnarray}%
which implies 
\begin{equation}
\Sigma _{k}=\frac{\Sigma _{\mathrm{d}}(k)}{1+G_{\mathrm{loc}}(i\nu
_{n})\Sigma _{\mathrm{d}}(k)}+\Sigma _{\mathrm{loc}}(i\nu _{n}).
\label{relSigma}
\end{equation}%
The result (\ref{relSigma}) was derived in Ref. \cite{DF1}.

\section{The effect of the six-point vertex}

\subsection{Relation between the dual and lattice self-energy}

The relation (\ref{relSigma}) does not change its form in the approximation,
when one neglects six-point (and higher order) local vertices in Eq. (\ref%
{VDMFT}). This however does not necessarily mean, that it remains correct in
this case. Instead, as becomes clear from the following discussion, the
relation (\ref{relSigma}) implicitly assumes that the one-particle reducible
diagrams for six- and higher order vertices are taken into account.

Let us consider the diagrams for the self-energy, which include six-point
one particle reducible vertex, such as the diagram shown in Fig. 1a. These
diagrams, being 1PI in terms of the vertices $V_{\mathrm{loc}}^{(6)}$,
produce nevertheless one-particle reducible contributions to the
self-energy, which should be excluded. The denominator in the relation (\ref%
{relSigma}) aims to remove the corresponding one-particle reducible diagrams
for self-energy. Formulated differently, the quantity $\Sigma ^{\prime }$ in
Eq. (\ref{relSigma}) must contain one-particle reducible diagrams, which are
cancelled by the denominator in the first term in Eq. (\ref{relSigma}).

\begin{figure}[tbp]
\includegraphics{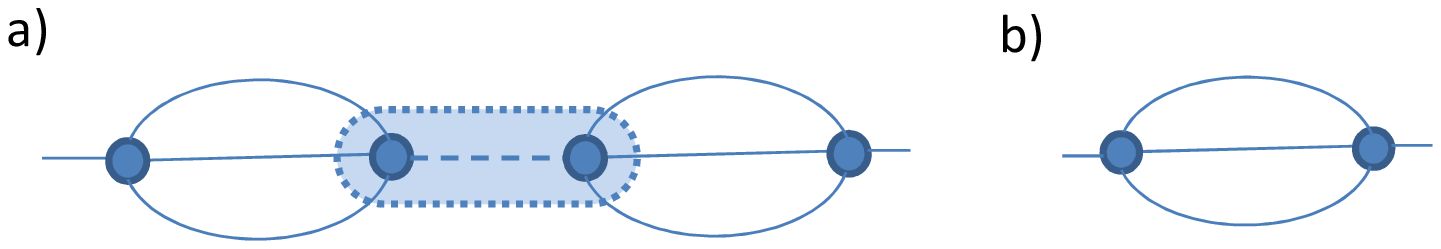}
\caption{(Color online) (a) The one-particle reducible contributions to the
lattice two-point Green function, coming from one-particle reducible
six-point local vertex (shaded). The circles correspond to the four-point
local vertices, solid line - the propagator $\widetilde{G}_{k}$, dashed line
corresponds to the local propagator $G_{loc}$. (b) The corresponding
contribution to the non-local self-energy, which is of the second order in
the local 4-point vertices.}
\end{figure}

To prove this statement for the diagrams, similar to that of Fig. 1a,
containing repeating lowest (second-order) diagram of Fig. 1b, it is
sufficient to consider the tree diagram contribution to the six-point vertex
of the form $V_{\mathrm{loc}}^{(6)}=\sum \Gamma _{\mathrm{loc}}G_{\mathrm{loc%
}}\Gamma _{\mathrm{loc}}$, the sum is taken over different combinations of $%
4 $-momenta, assigned to 4-point vertices. 
%These combinations yield however equalent contributions to the functional $%
%\mathcal{V}.$
Decoupling the resulting six-particle interaction with the fermionic
Hubbard-Stratonovich transformation by introducing auxiliary fermionic field 
$\phi _{k},$ we obtain%
\begin{eqnarray}
e^{-\widehat{\mathcal{V}}[\widehat{\eta },\widehat{\eta }^{+},\widehat{\zeta 
},\widehat{\zeta }^{+}]} &=&\int d[\widetilde{c},\widetilde{c}^{+}]\exp
\left\{ \sum\limits_{k,\sigma }\left[ \widetilde{G}_{k}^{-1}\left( 
\widetilde{c}_{k,\sigma }^{+}-\widehat{\eta }_{k,\sigma }^{+}\right) \left( 
\widetilde{c}_{k,\sigma }-\widehat{\eta }_{k,\sigma }\right) \right. \right.
\nonumber \\
&&\left. +G_{\mathrm{loc}}^{-1}(i\nu _{n})(\phi _{k,\sigma }^{+}-\widehat{%
\zeta }_{k,\sigma }^{+})(\phi _{k,\sigma }-\widehat{\zeta }_{k,\sigma })%
\right]  \label{Vez} \\
&&\left. -\widehat{\mathcal{V}}_{DMFT}^{\prime }[\widetilde{c}^{+},%
\widetilde{c},\phi ^{+},\phi ]\right\}  \nonumber \\
\widehat{\mathcal{V}}_{DMFT}^{\prime }[\widetilde{c}^{+},\widetilde{c},\phi
^{+},\phi ] &=&\frac{1}{2}\Gamma _{\mathrm{loc}}\circ (\widetilde{c}%
_{k_{1},\sigma }^{+}\widetilde{c}_{k_{3},\sigma })(\widetilde{c}%
_{k_{2},\sigma ^{\prime }}^{+}\widetilde{c}_{k_{4},\sigma ^{\prime }}) 
\nonumber \\
&&+\Gamma _{\mathrm{loc}}\circ (\widetilde{c}_{k_{1},\sigma }^{+}\widetilde{c%
}_{k_{3},\sigma })(\widetilde{c}_{k_{2},\sigma ^{\prime }}^{+}\phi
_{k_{4},\sigma ^{\prime }})  \nonumber \\
&&+\Gamma _{\mathrm{loc}}\circ (\widetilde{c}_{k_{1},\sigma }^{+}\widetilde{c%
}_{k_{3},\sigma })(\phi _{k_{2},\sigma ^{\prime }}^{+}\widetilde{c}%
_{k_{4},\sigma ^{\prime }})
\end{eqnarray}%
where we have introduced source fields $\widehat{\zeta }$ for fermions $\phi
.$ The effective interaction (\ref{Vez}) can be put in more compact form by
introducing spinors $\Phi _{k,\sigma }=(\widetilde{c}_{k,\sigma },\phi
_{k,\sigma })$ and $\Pi _{k\sigma }=(\widehat{\eta }_{k,\sigma },\widehat{%
\zeta }_{k,\sigma }),$ such that 
\begin{equation}
e^{-\widehat{\mathcal{V}}[\Pi ,\Pi ^{+}]}=\int d[\Phi ,\Phi
^{+}]e^{\sum\limits_{k,\sigma }\left( \Phi _{k\sigma }^{+}-\Pi _{k\sigma
}^{+}\right) \widehat{G}_{k}^{-1}\left( \Phi _{k\sigma }-\Pi _{k\sigma
}\right) -\widehat{\mathcal{V}}_{DMFT}^{\prime }[\Phi ^{+},\Phi ]}
\end{equation}%
where the corresponding matrix bare Green function reads 
\begin{equation}
\widehat{G}_{k}=\left( 
\begin{array}{cc}
\widetilde{G}_{k} & 0 \\ 
0 & G_{\mathrm{loc}}(i\nu _{n})%
\end{array}%
\right)  \label{Gt}
\end{equation}%
It is of crucial importance that the matrix Green function, Eq. (\ref{Gt}),
contains both, the non-local $\widetilde{G}_{k}$ and the local $G_{\mathrm{%
loc}}(i\omega _{n})$ components, which are mixed through the non-local dual
self-energy, as considered below.

The resulting two-point vertices can be also considered as matrices in the
space ($\widehat{\eta },\widehat{\zeta }$). The relation between the lattice
and dual two-point vertices then has the form, similar to the Eq. (\ref%
{Vdtol}),%
\begin{eqnarray}
V_{k}^{(2)} &=&\left( 
\begin{array}{cc}
\frac{\Sigma _{\mathrm{loc}}(i\nu _{n})}{1-\Sigma _{\mathrm{loc}}(i\nu
_{n})G_{0k}} & 0 \\ 
0 & 0%
\end{array}%
\right) +\widehat{V}_{k}^{(2)} \\
\widehat{V}_{k}^{(2)} &=&\widehat{\Sigma }_{\mathrm{d}}(k)[1-\widehat{G}_{k}%
\widehat{\Sigma }_{\mathrm{d}}(k)]^{-1},  \nonumber
\end{eqnarray}%
where $\widehat{\Sigma }_{\mathrm{d}}(k)$ is the self-energy matrix of $%
\widetilde{c}_{k,\sigma }$ and $\phi _{k,\sigma }$ fields, having both,
diagonal and off-diagonal contributions. For the second-order diagram of
Fig. 1b the self-energy is equal for both fermion species: $\ $%
\begin{equation}
\widehat{\Sigma }_{\mathrm{d}}(k)=\Sigma _{\mathrm{d}}^{1\mathrm{PI}%
}(k)\left( 
\begin{array}{cc}
1 & 1 \\ 
1 & 1%
\end{array}%
\right) .
\end{equation}%
where $\Sigma _{\mathrm{d}}^{1\mathrm{PI}}(k)$ is the value of the 1PI
diagram of Fig. 1b. From this we obtain the relation between the local and
lattice self-energies%
\begin{eqnarray}
V_{k}^{(2)\widetilde{c},\widetilde{c}} &=&\frac{\Sigma _{\mathrm{loc}}(i\nu
_{n})}{1-\Sigma _{\mathrm{loc}}(i\nu _{n})G_{0k}}+\frac{\Sigma _{\mathrm{d}%
}^{1\mathrm{PI}}(k)}{1-G_{k}\Sigma _{\mathrm{d}}^{1\mathrm{PI}}(k)}\frac{1}{%
\left[ 1-\Sigma _{\mathrm{loc}}(i\nu _{n})G_{0k}\right] ^{2}}  \nonumber \\
&=&\frac{\Sigma _{k}}{1-\Sigma _{k}G_{0k}},
\end{eqnarray}%
which yields%
\begin{equation}
\Sigma _{k}=\Sigma _{\mathrm{loc}}(i\nu _{n})+\Sigma _{\mathrm{d}}^{1\mathrm{%
PI}}(k).  \label{Sigma2}
\end{equation}%
The result (\ref{Sigma2}) is essentially different from Eq. (\ref{relSigma})
and implies that the one-particle reducible contributions in the
self-energy, occurring due to the one-particle reducible contributions to
the six-point vertex, are indeed cancelled by the denominator in Eq. (\ref%
{relSigma}). Equtions (\ref{relSigma}) and (\ref{Sigma2}) also imply the
analogue of the Dyson equation for the dual fermion self-energy 
\begin{equation}
\Sigma _{\mathrm{d}}(k)=\frac{\Sigma _{\mathrm{d}}^{1\mathrm{PI}}(k)}{1-G_{%
\mathrm{loc}}(i\nu _{n})\Sigma _{\mathrm{d}}^{1\mathrm{PI}}(k)}.
\label{Sigma_rel1PI}
\end{equation}%
Having the 1PI self-energy $\Sigma _{\mathrm{d}}^{1\mathrm{PI}}(k)$ the
result (\ref{Sigma2}) should be used to obtain the lattice self-energy
instead of the equation (\ref{relSigma}), suggested by Refs. \cite%
{DF1,DF2,DF3,DF4}.

\begin{figure}[tbp]
\includegraphics{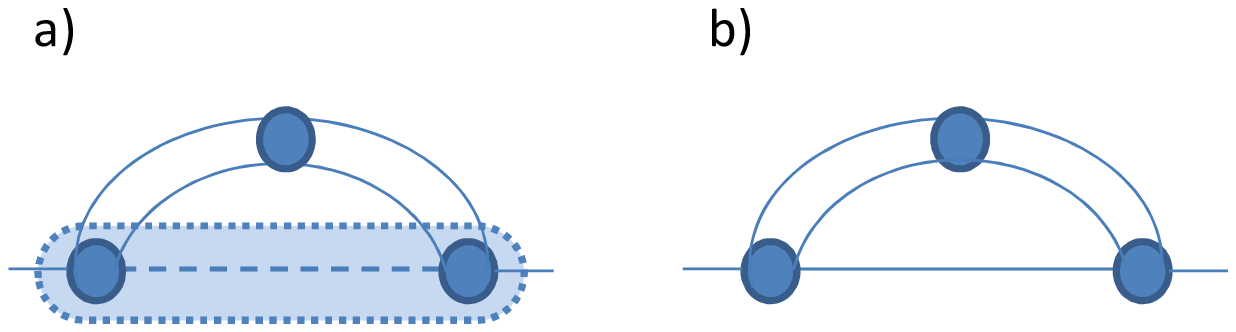}
\caption{(Color online) The non-local contributions to the lattice
self-energy, which are of the third order in four-point local vertices. (a)
The contribution, containing the local Green function, coming from the
one-particle reducible six-point vertex. (b) Similar contribution of the
dual fermion approach, neglecting six-point vertices. Notations are the
same, as in Fig. 1}
\end{figure}

For the considered theory with only six-point one-particle reducible
contributions included, the newly derived relation (\ref{Sigma2}) between
the dual and lattice self-energy is fulfilled if (and only if) the
self-energy is equal for both fermion species $\widetilde{c}$ and $\phi ,$
as it happens for the diagram of Fig. 1b. The abovementioned assumption does
not necessarily hold in higher orders of dual perturbation theory. However,
the inclusion of one-particle reducible contributions of higher order (eight
and more point vertices) makes the relations (\ref{Sigma2}) and (\ref%
{Sigma_rel1PI}) fulfilled in more general situations\cite{OurCommon}.

The one-particle reducible contributions to the six-point vertex can also
produce 1PI self-energy diagrams, containing local Green functions, such as
shown in Fig. 2a. These diagrams, being formally of the same order in the
local 4-point vertices, as considered by the dual fermion approach (Fig.
2b), are not taken into account when the six-point and higher vertices are
not taken into account. At the same time, the diagrams, similar to that
shown in Fig. 2a, can produce even larger contribution to the self-energy,
than the diagrams of the dual fermion approach, neglecting 6-point and
higher vertices, since the sum of the local Green function over momentum
does not vanish. Therefore, accounting one-particle reducible parts of six-
and higher-order local vertices appears to be crucially important for both,
the diagrammatic consistency of the dual fermion approach and keeping all
the diagrams of the same order in the four-point local vertices.

We also note that the covariation splitting method, used in the dual fermion
approach (\ref{Vs1}), is similar to that applied in functional
renormalization-group approach (see, e.g., Refs. \cite{Polchinskii,Salmhofer}%
), except that the latter considers integration of degrees of freedom in
many infinitesimally small steps, while the former -- only in two steps.
Similarly to the discussion above, in the Polchinski formulation of the
functional renormalization-group approach \cite{Polchinskii,Salmhofer}
one-particle reducible contributions to six-point vertices were argued to be
important for proper calculation of the four-point vertices already in
one-loop approximation \cite{Zanchi}. The same contributions can be also
shown to be important for evaluation of multi-loop contributions to the
self-energy. In the next subsection we address another aspect, where the
six- and higher-order vertices appear to be important in the dual fermion
approach.

\subsection{Self-energy feedback}

The dual fermion approach accounts for the self-energy feedback by dressing
the Green function of the non-local degrees of freedom:%
\begin{equation}
\widetilde{\mathbb{G}}_{k}=\widetilde{G}_{k}/[1-\widetilde{G}_{k}\Sigma _{%
\mathrm{d}}(k)].  \label{GD}
\end{equation}%
However, the function $\widetilde{G}_{k}$, which represents a difference of
two propagators, does not correspond to a physically observable quantity,
and it is informative to trace, how dressing it one can finally obtain the
physical propagator%
\begin{eqnarray}
\mathbb{G}_{k} &=&G_{k}/[1-G_{k}(\Sigma _{k}-\Sigma _{\mathrm{loc}})]
\label{GDD} \\
&=&G_{k}/[1-G_{k}\Sigma _{\mathrm{d}}^{1\mathrm{PI}}(k)],  \nonumber
\end{eqnarray}%
which is constructed by dressing the Green function $G_{k},$ see Eq. (\ref%
{Glattice}), containing only the local self-energy, by the remaining
self-energy difference $\Sigma _{k}-\Sigma _{\mathrm{loc}}.$ We have
observed in Sect. 3.1, that in the lowest orders of perturbation theory,
propagators $\widetilde{\mathbb{G}}_{k},$ appearing in the diagram technique
for Eq. (\ref{Vez1}), are added by $G_{\mathrm{loc}},$ coming either from
either adding local quantities to their non-local counterparts (such as in
Eq. (\ref{Sigma2})), or from the contributions, containing one-particle
reducible six-point and higher order vertices, such as the diagram of Fig.
2a. Adding $G_{\mathrm{loc}}$ to $\widetilde{\mathbb{G}}_{k}$ is however
still not sufficient to reproduce (\ref{GDD}) for $\Sigma _{\mathrm{d}}^{1%
\mathrm{PI}}(k)\neq 0.$

Again, we argue, that the six- and higher order local vertices are crucially
important to obtain (\ref{GDD}). To see this, let us insert $G_{k}=%
\widetilde{G}_{k}+G_{\mathrm{loc}}$ into Eq.(\ref{GDD}), use Eqs. (\ref%
{Sigma_rel1PI}) and (\ref{GD}) to represent the result in terms of $%
\widetilde{\mathbb{G}}_{k}$ and $\Sigma _{\mathrm{d}}^{1\mathrm{PI}}(k)$,
and expand the result in a series of $\Sigma _{\mathrm{d}}^{1\mathrm{PI}%
}(k): $%
\begin{equation}
\mathbb{G}_{k}=\sum_{n=1}^{\infty }[G_{\mathrm{loc}}(i\nu _{n})+n\widetilde{%
\mathbb{G}}_{k}][G_{\mathrm{loc}}(i\nu _{n})\Sigma _{\mathrm{d}}^{1\mathrm{PI%
}}(k)]^{n-1}
\end{equation}

\begin{figure}[tbp]
\includegraphics{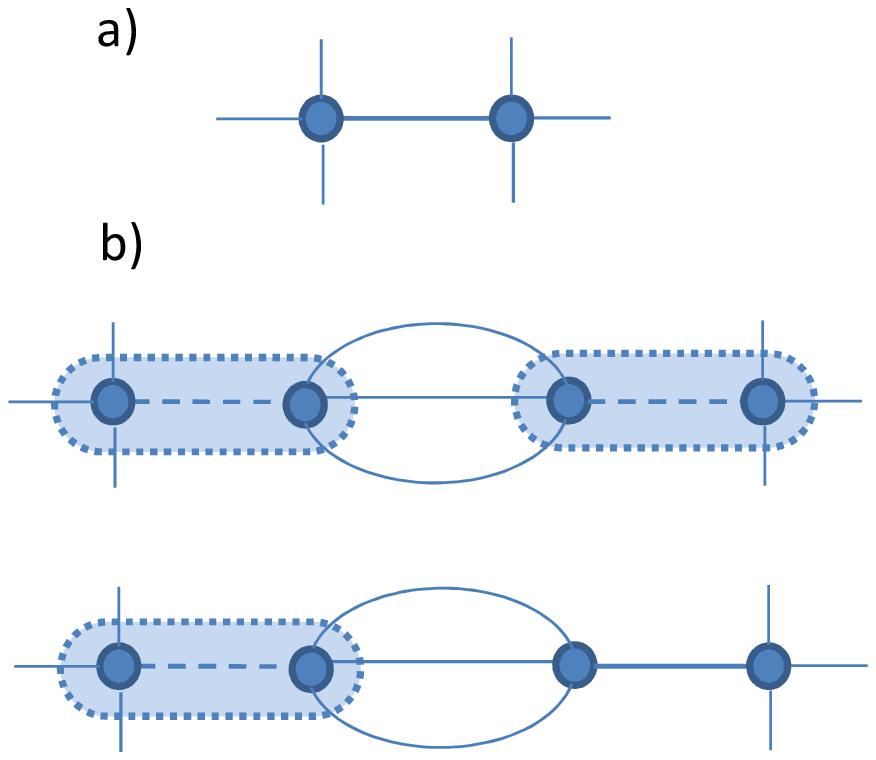}
\caption{(Color online) Dressing of the non-local Green function $\widetilde{%
\mathrm{G}}_{k}$, shown by bold line and appearing as a part of diagram (a)
by the two contributions of the lowest-order in $\Sigma ^{1\mathrm{PI}}_{%
\mathrm{d}} (k)$ (b). Notations are the same, as in Fig. 1}
\end{figure}

The term with $n=1$ represents the combination $G_{\mathrm{loc}}+\widetilde{%
\mathbb{G}}_{k},$ discussed above, while the terms with $n\geq 2$ in this
series expansion can be ascribed to the respective diagrams (see Fig. 3 for $%
n=2$), where the one-particle irreducible local vertices are connected by
local propagators, forming one-particle reducible six-point and higher
vertices. The dual fermion approach, which does not account for the
six-point and higher order vertices, neglects therefore a difference between 
$\mathbb{G}_{k}$ and $\widetilde{\mathbb{G}}_{k}+G_{\mathrm{loc}}.$

\section{Conclusion}

In the present paper we have considered effect of one-particle reducible
six- and higher-point vertices in the dual fermion approach. We have argued
that the one-particle reducible contributions to these vertices are
important to make the dual fermion approach diagrammatically consistent.
Neglecting the six-point and higher order vertices does not allow to obtain
correct relation between the dual and lattice self-energies, as well as
treat correctly the feedback of the dual self-energy on the dual Green
functions. Apart from that, the one-particle reducible six-point and higher
order vertices lead to the self-energy corrections, 
%, which are of the same order
%in the four-point vertices, as those, retained in the dual fermion approach,
%neglecting six-point and higher order vertices, but
which contain both, local and non-local Green functions.

Further numerical investigations of the (un)importance of the described
contributions of six-point and higher-order vertices to be performed (see,
e.g., Ref. \cite{OurCommon}). This also calls for developing a new method of
treatment of non-local degrees of freedom, which avoids the described
problems of the dual fermion approach and operates with the one-particle
irreducible quantities.

\textit{Acknowledgements. }The author is grateful to G. Rohringer, A.
Toschi, K. Held, C. Honerkamp, and W. Metzner for stimulating discussions.
The work is supported by the RFBR grant 10-02-91003-ANFa.

\end{document}